\documentclass[aps,prl,preprint,showpacs]{revtex4}
\usepackage{mathrsfs}
\usepackage{graphicx}
\usepackage{amssymb}
\usepackage{amsmath}
\begin{document}

\title{Spontaneous emission of a nanoscopic emitter in a strongly scattering disordered medium}

\author{P. V. Ruijgrok,$^{1*}$ R. W\"{u}est,$^1\dag$ A. A. Rebane$^{1}$, A. Renn$^{1}$, and \\V. Sandoghdar$^{1\ddag}$}

\affiliation{
$^{1}$Laboratory of Physical Chemistry and optETH, ETH Z\"{u}rich, CH-8093 Z\"{u}rich,
Switzerland\\
$^*$ Present address: Huygens Laboratory, Leiden University, 2300 RA Leiden, The Netherlands\\
$^\dag$ Present address: ABB Ltd, Corporate Research, CH-5405 Baden-D\"attwil, Switzerland\\
$^\ddag$ vahid.sandoghdar@ethz.ch\\
http://www.nano-optics.ethz.ch}

\begin{abstract}
Fluorescence lifetimes of nitrogen-vacancy color centers in
individual diamond nanocrystals were measured at the interface
between a glass substrate and a strongly scattering medium.
Comparison of the results with values recorded from the same
nanocrystals at the glass-air interface revealed fluctuations of
fluorescence lifetimes in the scattering medium. After discussing a
range of possible systematic effects, we attribute the observed
lengthening of the lifetimes to the reduction of the local density
of states. Our approach is very promising for exploring the strong
three-dimensional localization of light directly on the microscopic
scale.\end{abstract}

\maketitle

The behavior of electromagnetic waves in disordered media has been a
subject of intense studies in the past
decade~\cite{Wiersma1997,Chabanov2000,Storzer2006,Schwartz:07,Barthelemy:08,Aegerter2009}.
In particular, the regime of strong scattering where the mean free
path of a photon becomes comparable to its wavelength has fascinated
scientists. In such systems, constructive and destructive
interference can lead to the modification of the local density of
states (LDOS) in various spatial modes and to the localization of
light~\cite{Mirlin2000,Schomerus2002,FroufePerez2007}. In an ideal
``gedanken" experiment, one might imagine to map the fluctuations of
LDOS directly at the subwavelength level. However, so far, light
scattering phenomena have been studied via transmission and
reflection intensity measurements on macroscopic samples, where
separation of the effects of absorption and scattering has posed a
challenge~\cite{Wiersma1997,Aegerter2009,Storzer2006,
Chabanov2000,Hu2008}. In this work, we probe the variations of LDOS
by measuring the modification of the spontaneous emission rate of
individual nanoscopic emitters. Our experimental strategy is to
characterize the fluorescence decay of each emitter first and then
trace its change after covering them by a scattering disordered
medium.

As emitters, we used nitrogen-vacancy (NV) color centers in diamond
nanocrystals (DNC). The size of the DNCs ranged between 10 and 500
nm with an average size of 60 nm, as measured by electron
microscopy. The DNCs were treated by irradiation with 1.5 MeV
electrons during 8 hours with a total dose of 3 $10^{18}$ e/cm$^2$.
Annealing at 850 C for 4 hours resulted in approximately 1-10 NV
color centers in each DNC. Aside from having a high quantum yield,
our choice of emitter has two crucial advantages. First, NV centers
in diamond are indefinitely photostable so that repeated
quantitative measurements can be performed. Second, the color
centers are well protected in the diamond lattice against surface
effects upon contact with other material~\cite{Tisler2009}.

Our scattering sample consisted of a powder of rutile $\rm TiO_2$
particles (DuPont, Ti-Pure R706) with a mean size of 250~nm and a
distribution of 150-350 nm as revealed by electron microscopy.
Particles were coated with silica (SiO2, 3~wt~\%) and alumina ($\rm
Al_2O_3$ 2.5~wt~\%) and had a refractive index of
2.8~\cite{Storzer2006}. We produced a disordered medium in half
space by sprinkling the powder on a cover glass carrying the DNCs.
We then compressed the powder manually by applying a gentle pressure
via another glass cover slide. The final thickness of the medium was
typically about 0.4~mm. Based on the volume and weight of the
material that was deposited, we estimated the volume fraction
occupied by particles in the compressed powder to be about 30\%.

\begin{figure}[tb]
\centerline{\includegraphics[width=9.5cm]{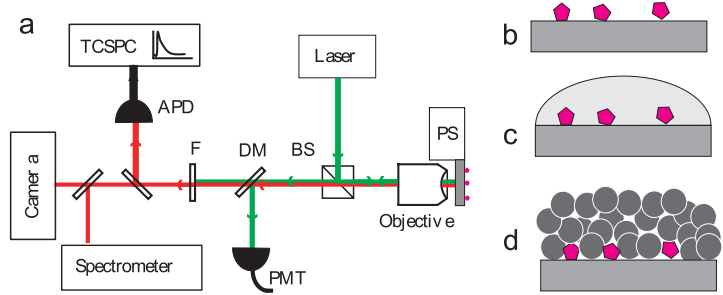}}
\caption{a) Schematics of the experimental setup. BS, 50/50 beam
splitter; DM, dichroic mirror; PS, piezoelectric scanner; PMT,
photomultiplier tube; F, long pass filter; APD, avalanche
photodiode; TCSPC, time correlated single photon counting card. b-d)
Sample arrangements for diamond nanocrystals exposed to air, water,
and $\rm TiO_2$ particles. }\label{setup}
\end{figure}

The schematics of the experiment are shown in Fig.~\ref{setup}. The
sample was placed on a three-dimensional piezoelectric stage of a
home-built microscope. Laser pulses at a wavelength of 532 nm,
duration of 13~ps, and repetition rate of 3.8 MHz were used to
excite the NV centers. This light was sent to a microscope objective
(NA=1.4) via a 50/50 beam splitter. The scattering and fluorescence
from the sample were collected through the same objective, separated
via a dichroic mirror and two 540~nm long pass filters, and directed
to a photomultiplier tube and an avalanche photodiode, respectively.
Fluorescence lifetimes were determined with a time-correlated
single-photon counter by recording the time delay between the
excitation pulses and the detected photons. The intensity $I$
measured on the photomultiplier tube resulted from the interference
of the light reflected by the sample interface and that scattered by
a given nano-object in the focus spot. The details of this
interferometric scattering (iSCAT) detection method are described in
our previous work~\cite{Lindfors:04,Kukura:09}. The added value of
the iSCAT signal is to visualize nano-objects without the need for a
fluorescence signal.

\begin{figure}[b]
\label{fig:images}
\centerline{\includegraphics[width=8.5cm]{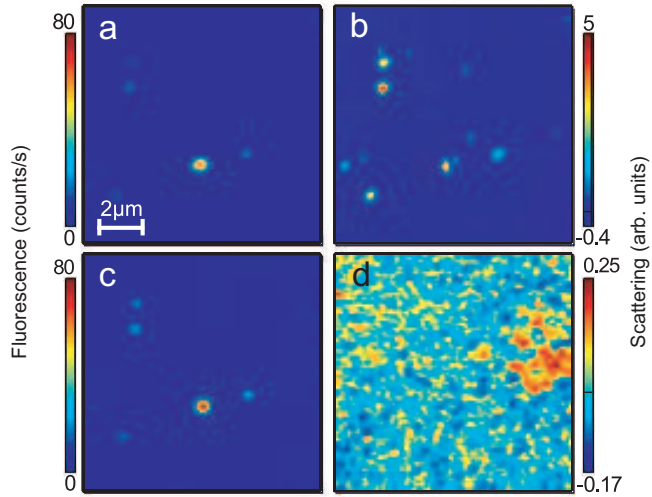}}
\caption{Fluorescence (a,c) and iSCAT (b,d) images of the same
diamond nanoparticles on a glass cover slide in air (a,b) and
covered with scattering particles (c,d).} \label{images}
\end{figure}

In a first step, the DNCs were spin coated on a glass cover slide
(see Fig.~\ref{setup}b) and simultaneously imaged in fluorescence
and iSCAT by scanning the sample across the focus of the excitation
laser beam. Figure~\ref{images}a shows an example of a fluorescence
map of DNCs while Fig.~\ref{images}b displays the simultaneously
recorded iSCAT image. The latter locates the same DNCs as well as a
few other nonfluorescent nanoparticles which are most likely DNCs
without color centers.

\begin{figure}[tb]
\centerline{\includegraphics[width=11cm]{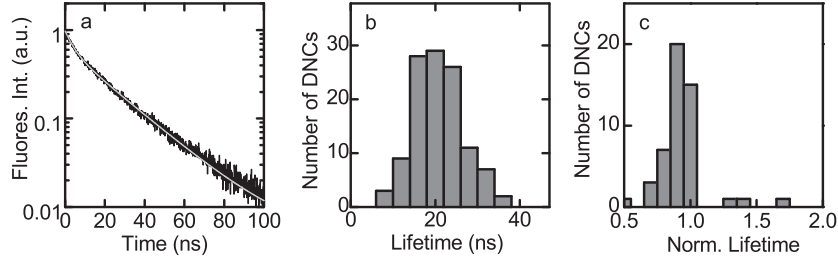}} \caption{a)
Fluorescence decay curve measured from a single nanoparticle. b)
Histograms of fluorescence lifetimes at the air-glass interface. c)
The spread of the ratio $\tau_{\rm w} / \tau_{\rm a}$ between the
measured lifetimes at the water-glass and air-glass
interfaces.}\label{curves-1}
\end{figure}

The second step of the experiment involved the measurement of the
excited state lifetime $\tau$ for each fluorescent DNC. For NV
centers in bulk diamond, $\tau$ has been measured to be about
12~ns~\cite{Collins1983}. It is known, however, that the spontaneous
emission rate is slowed down, and thus $\tau$ becomes larger in
subwavelength dielectric media~\cite{schniepp:02}. Furthermore, as
the size of the dielectric medium is decreased, the dependence of
the spontaneous emission rate on the shape of the medium as well as
the emitter orientation and position becomes less
significant~\cite{Rogobete2003}. Figure~\ref{curves-1}a displays a
typical example of measurements on a single DNC. This measurement is
fitted using a background and a double exponential with $1/e$ times
of $4.2 \pm 0.3$~ns and $22.8\pm 0.3$~ns, respectively. We attribute
the deviation from a simple exponential decay to the inhomogeneity
of emitting transitions in a single DNC. In particular, the shorter
lifetime component is most likely due to quenching effects for some
NV centers. A plausible cause of this might be the presence of a
thin graphite layer on the DNCs~\cite{Tisler2009}. Another source of
inhomogeneity can stem from the coexistence of a transition at
575~nm, which is attributed to the neutral charge state of NV
centers, and the usual transition of the negatively-charged center
at 637 nm~\cite{Jelezko2006}. Indeed, we confirmed the presence of
both lines by measuring the emission spectra of individual DNCs. To
our knowledge, no data are available on the fluorescence lifetime of
the neutral color centers. A third phenomenon that can cause
different fluorescence lifetimes in a DNC is the orientation of the
transition dipole moments of the individual NV centers with respect
to the glass substrate~\cite{Barnes:98}. To examine this effect, we
have performed rigorous finite-difference time-domain (FDTD)
calculations to compute the variation of the radiative decay rates
of emitters in a dielectric nano-object placed on a glass
surface~\cite{Chen:10}. It turns out that for the parameters
(indices of refraction, particle size) of relevance to our
measurements, $\tau$ could vary within a factor of three, depending
on the orientation of each emitter and its distance from the
interface.

The details of the inhomogeneities in $\tau$ are not important to
our current work because we will be only interested in its
modification when the scattering medium is added. In fact, in what
follows we will only investigate the changes of the long-lifetime
component of the measurements. As explained below, while several
systematic effects could cause shortening of the fluorescence
lifetime, its increase provides a robust evidence for the
modification of the density of states in the scattering medium.
Fig.~\ref{curves-1}b displays the histogram of this component for
DNCs placed on a glass cover slide in air. The variation from 8 ns
to 37 ns is in agreement with previous
reports~\cite{Beveratos2001,Tisler2009}. We checked the validity of
this attribution by measuring the fluorescence lifetime $\tau_{\rm
w}$ of the same DNCs after adding water on top of the sample (see
Fig.~\ref{setup}c). The change in the radiative lifetime of an
emitter is inversely proportional to the refractive index of the
environment so we expect $\tau$ to be reduced under water.
Figure~\ref{curves-1}c shows a histogram of the resulting values
normalized to the lifetimes of each DNC in air. The mean value of
$\tau_{\rm w}/\tau_{\rm a}$ amounts to 0.9, where $\tau_{\rm a}$
denotes the fluorescence lifetime at the air-glass interface. This
outcome is in very good agreement with the expected value of 0.88 if
one considers the refractive index of the surrounding environment to
be the average of the refractive indices of the media in the lower
and upper halves of the glass-air or glass-water interfaces. Note
that we do not have to account for local field effects since the NV
centers embedded in the diamond matrix do not experience a change of
environment at the molecular level~\cite{schniepp:02}.

\begin{figure}[tb]
\centerline{\includegraphics[width=7.5cm]{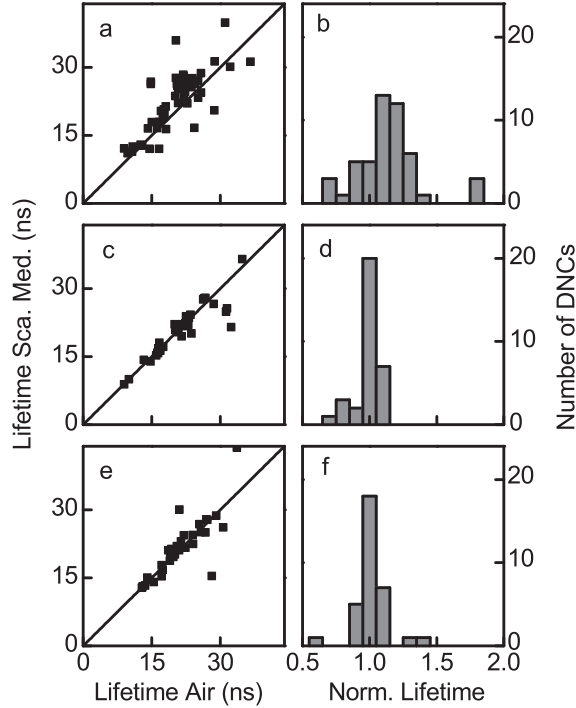}}
\caption{a-f) Data collected on the fluorescence lifetimes of many
individual DNCs in three different samples (First row, sample 1;
second row, sample 2; third row, sample 3.) measured in air and in
the scattering medium. We examined 49, 33, and 33 DNCs in the three
studies, respectively.}\label{curves-2}
\end{figure}

Once the DNCs were characterized, the water was removed with a
micro-pipette and the substrate was left to dry overnight under a
gentle stream of nitrogen. In the last phase of the experiment, $\rm
TiO_2$ particles were placed on the sample (see Fig.~\ref{setup}d).
Figures~\ref{images}c and d show the fluorescence and iSCAT images
in the scattering medium. The individual DNCs are still easily
identified via fluorescence, while the scattering image now displays
a ``speckle" pattern. Figures~\ref{curves-2}a-f present the measured
lifetimes of many DNCs from three different scattering samples (the
step with water was omitted for samples 2 and 3). Let us label the
fluorescence lifetime in the scattering medium by $\tau_{\rm s}$. As
in the case of water immersion, we found a distribution of the
normalized fluorescence lifetimes $\tau_{\rm s}/\tau_{\rm a}$,
indicating both reduction and enhancement of the spontaneous
emission rate. However, while in the previous case the average value
$\langle \tau_{\rm w}/ \tau_{\rm a} \rangle<1$, here we find
$\langle \tau_{\rm s}/ \tau_{\rm a} \rangle>1$.  In sample 1, for
example, fluorescence lifetimes are longer than in air by up to a
factor of 1.4 with an ensemble average of 1.1. The histograms of
Figs.~\ref{curves-2}b, d, f reveal variations in the measurements
from the three samples. A possible cause of this might be different
qualities of powder compression and nanoparticle packing.

The observation of lifetimes longer than in air indicates the
reduction of LDOS in the scattering medium. We believe this
conclusion is robust against systematic effects. First, opening of
non-radiative decay channels, e.g. as a result of contact with the
medium, cannot be responsible for this effect since it could only
increase the total decay rate $\gamma=\gamma_{\rm r}+\gamma_{\rm
nr}$ and therefore lower the lifetime. Second, lengthening of the
fluorescence lifetime cannot be explained in the context of a
homogeneous medium with an effective index of refraction because
addition of high-index $\rm TiO_2$ particles could only increase
such an effective refractive index with respect to that of the
air-glass interface, and thus, reduce the lifetime. Finally, we
verified that the longer fluorescence decay was not the result of
photon trapping in the scattering medium. To do this, we focused
10~ps pulses of laser light onto the interface between the cover
glass and the scattering medium to mimic the spatial mode of the DNC
emission. We then interrogated the temporal profile of the reflected
light in a confocal detection arrangement with and without the
scattering medium. This allowed us to rule out any pulse lengthening
within the time resolution of 0.4~ns.

Our measurements provide the first report of the fluctuations of the
spontaneous emission rate in a disordered scattering medium. Our
experimental approach sets the ground for studying strong
localization of light by monitoring and mapping the variations of
the local density of states directly in the coordinate space. Such
studies promise to connect the nanoscopic behavior of light
scattering with conventional macroscopic ensemble measurements. An
important advantage of this approach is that observation of the
inhibition of the spontaneous emission rules out systematic effects
of absorption since those would always lead to a reduction of the
fluorescence lifetime.

The experiments reported here can be improved in several ways.
First, particle composition and compactification of the scattering
medium can be varied and characterized~\cite{Popescu:99}. Second,
use of color centers with narrow emission spectrum in the near
infrared~\cite{Gaebel2004} could simplify the analysis of the
density of states in comparison to the broad emission of NV centers.
Another advantage of such emitters would be the availability of very
strongly scattering media based on large-bandgap
semiconductors~\cite{Rivas2002}. Furthermore, we plan to extend our
work to the measurement of the fluorescence lifetime via a thin
optical fiber that carries a single DNC attached to its end placed
deep into a three-dimensional scattering medium~\cite{Kuehn:01}.
Finally, we hope that theoretical progress in modeling the
modification of radiative effects in strongly scattering media will
allow quantitative comparisons with experimental data.

%\textbf{Acknowledgements}\\
We thank H. Eghlidi for help with electron microscopy, P. Kukura for
help with the scattering detection and the data acquisition
software, and X. Chen for the FDTD calculations. V.S. acknowledges
illuminating discussions with R. Carminati and A. Dogariu.

\bibliography{vahid_edit}

\end{document}